\documentclass[epj,  nopacs]{svjour}
\usepackage{amsmath,amssymb,graphicx,bm}
\usepackage{color}
\usepackage{multicol}        
\usepackage[bottom]{footmisc}

\newcommand{\beq}{\begin{equation}}
\newcommand{\eeq}{\end{equation}}
\newcommand{\bea}{\begin{eqnarray}}
\newcommand{\eea}{\end{eqnarray}}
\newcommand{\ba}{\begin{align}}
\newcommand{\ea}{\end{align}}
\newcommand{\bfig}{\begin{figure}}
\newcommand{\efig}{\end{figure}}

\newcommand{\D}{\displaystyle}

\newcommand{\gev}{\, \text{GeV}}
\newcommand{\mev}{\, \text{MeV}}

\newcommand{\tin}{t_{\rm in}}
\newcommand{\la}{\langle}
\newcommand{\ra}{\rangle}

\newcommand{\tplus}{t_{+}}
\newcommand{\kpi}{K\pi}
\newcommand{\omnes}{{\cal{O}}}
\newcommand{\rsqpiks}{\la r^2_{\pi\text{K}} \ra}

\newcommand{\ct}{\Delta_{K\pi}}
\newcommand{\ctbar}{\bar\Delta_{K\pi}}

\begin{document}


%
\title{Stringent constraints on  the scalar $\kpi$  form factor from analyticity,  unitarity  and low-energy theorems}
\author{Gauhar Abbas$^{a}$  \and 
B.\ Ananthanarayan$^{a, *}$  \and 
I. Caprini$^{b}$  \and I. Sentitemsu Imsong$^{a}$  \and S. Ramanan$^{a}$}

\institute{$^a$ Centre for High Energy Physics,
Indian Institute of \\
Science, Bangalore 560 012, India \\
$^b$ National Institute of Physics and Nuclear Engineering,\\
Bucharest, R-077125, Romania
                                                       \\
{\email{$^*$anant@cts.iisc.ernet.in}}
}        

\date{\today}

\abstract{ We investigate  the scalar $\kpi$ form factor at low energies by the method of unitarity bounds  adapted  so as to include information
 on the phase and  modulus  along the elastic region  of the unitarity cut.  Using at input the values of the form factor at $t=0$ and the Callan-Treiman point, we obtain stringent constraints on the slope and curvature parameters of the Taylor expansion at the origin. Also, we predict a quite narrow  range for the higher order ChPT  corrections  at the second Callan-Treiman point.}

\titlerunning{K$\pi$ constraints and low-energy theorems}

\authorrunning{Gauhar Abbas et al.} 

\maketitle

\section{Introduction}
\label{introduction}

The low energy properties of the $\kpi$  form factors are of great interest both  experimentally and theoretically. In particular, a precise knowledge of the slope and curvature parameters at $t=0$  would serve to improve the experimental analysis  of $K_{l3}$ decays, confirm the predictions of chiral perturbation theory (ChPT)  and provide benchmarks for future lattice determination of these quantities.

In the present paper we  consider the scalar $\kpi$ form factor,  expanded as
\beq
       f_{0}(t) = f_0(0)\left(1 +\lambda_0'\, \frac{t}{M_\pi^2} +  \frac{1}{2} \lambda_0''\, \frac{t^2}{M_\pi^4} + \cdots\right)
       \label{eq:f0},
\eeq
in the physical region of $K_{l3}$ decay.  The dimensionless parameters $\lambda_0'$ and $\lambda_0''$  are related   by  $\lambda_0'= M_\pi^2  \rsqpiks/6$ and  $\lambda_0''= 2 M_\pi^4 c$ to the  radius  $\rsqpiks$ and curvature $c$ used alternatively in the literature. 

The $\kpi$ form factors have been calculated at low energies in ChPT 
\cite{GaLe19851,GaLe19852,BiTa} and on the lattice (for recent reviews see \cite{LL2009,HL2009}). At $t=0$, the present value  $f_+(0)=0.962\pm 0.004 $ \cite{LL2009} shows that the corrections to the  Ademollo-Gatto theorem are quite small. 
Other low energy theorems frequently used are \cite{CallanTreiman}-\cite{Oehme}
\beq\label{eq:CT}
f_{0}(\Delta_{K \pi}) =\frac{F_K}{F_\pi} + \Delta_{CT}, \quad  f_{0}(\bar\Delta_{K\pi}) =\frac{F_\pi}{F_K} + \bar\Delta_{CT}, \quad 
\eeq
where $\Delta_{K\pi} = M^{2}_{K}-M^{2}_{\pi}$ and  $\bar\Delta_ {K\pi} = -\Delta_{K\pi}$ are the  first and second Callan-Treiman points, respectively. The lowest order values are known from $F_K/F_\pi =1.193 \pm 0.006$  \cite{LL2009}, and the corrections calculated to one loop are $\Delta_{CT}=-3.1\times 10^{-3}$  and  $\bar \Delta_{CT}=0.03$  \cite{GaLe19852}. The higher order corrections appear to be negligible at the first point  \cite{KaNe}, but are expected  to be quite
 large  at the second one.

 Analyticity and unitarity represent a powerful tool for obtaining  information on the $K\pi$ form factors. Several  comprehensive dispersive analyses were performed recently, using either the coupled channels  Muskhelishvili - Omn\`es equations \cite{Jaminscalar2002,Bachirscalar2009}, or a single channel Omn\`es representation  \cite{BePa}.  

Alternatively, the method of unitarity bounds, proposed a long time ago in
\cite{Okubo,SiRa},  and applied since then to various electromagnetic and weak form factors, exploits the fact that a bound on an integral of the modulus squared of the form factor along the unitarity cut is sometimes known  from independent sources.
Standard mathematical techniques then allow one to correlate the values of the form factor at different points or to control the truncation error of power expansions used in fitting the data \cite{BoCaLe}.

 For the $\kpi$ system the method was applied in \cite{BoMaRa} and more recently in \cite{BC,AbAn}. In Ref. \cite{BC} the method was extended by including the phase of the scalar form factor along the elastic part of the cut, known from the elastic $K\pi$ scattering by Watson's theorem, while in \cite{AbAn} information on the form factor at the second Calln-Treiman point was included for the first time in the frame of the standard bounds.

In the present work we revisit the issue of bounds on the expansion coefficients (\ref{eq:f0})  by applying a more sophisticated version of the unitarity
bounds  proposed  in \cite{Caprini2000}. The method 
uses the fact that the knowledge of the phase  allows one to remove the elastic cut  and define  a function with a larger analyticity domain. To be optimal, the method requires also some information on the modulus of the form factor in the elastic region.  In \cite{BC}, where the phase  constraint was treated using
Lagrange multipliers, this stronger property of the phase was not exploited, since no experimental information on the modulus was available at that time.

More recently, the precise measurements of the $\tau\to K\pi\nu_\tau$ spectral function by Belle collaboration \cite{Belle} provided also a first direct experimental determination of the modulus of the $K\pi$ form factors below a certain energy. The modulus is available also from the dispersive analyses  \cite{Jaminscalar2002,Bachirscalar2009}.  This justifies the application of the method proposed in  \cite{Caprini2000}. Our work extends the analysis made in \cite{AbAn} by including information  on the phase and modulus of the form factor
 on a part of the cut, which leads to a considerable improvement of  the bounds.  In the next section  we describe briefly the method for the scalar form factor, and in section \ref{sec:results} we present the results. A more detailed analysis, including  a discussion of the experimental implications and of  the vector form factor,  will be presented in  \cite{AACIR2}.

\section{Standard and new unitarity  bounds}\label{sec:bounds}
The method makes use of the following mathematical result: let $g(z)$ be a function analytic in the unit disk $|z|<1$ of the complex $z$-plane, which satisfies the inequality:
\beq	\label{eq:gI}
\frac{1}{2 \pi} \int^{2\pi}_{0} d \theta |g(\exp(i\theta))|^2 \leq I,
\eeq
where $I$ is a positive number. Then, if 
\beq\label{eq:gz}
g(z) = g_{0} + g_{1} z + g_{2} z^{2} + \cdots	
\eeq
is the Taylor expansion of $g(z)$ at $z=0$, and $g(z_1)$, $g(z_2)$ denote the values of $g$ at two points inside the analyticity domain, $|z_1|<1,~ |z_2|<1$ (for simplicity we assume that $z_1$ and $z_2$ are  real), the following determinantal inequality holds: 
\beq	\label{eq:det}
\left|
	\begin{array}{c c c c c c}
	I & g_0 & g_1 & g_2 & g(z_1) & g(z_2)\\	
	g_0 & 1 & 0 & 0 & 1 & 1\\
	g_1 & 0 & 1 & 0 & z_1 & z_2 \\
	g_2 & 0 & 0 & 1 & z_1^2 & z_2^2 \\
	g(z_1) & 1 & z_1 & z_1^2 & (1-z_1^2)^{-1} & (1-z_1 z_2)^{-1}\\
	g(z_2) & 1 & z_2 & z_2^2 & (1-z_2 z_1)^{-1} & (1-z_2^2)^{-1} \\
	\end{array}\right|\ge 0.
\eeq
Moreover, all the  principal minors of the above matrix should be nonnegative. 
For the proof see Refs. \cite{Okubo,SiRa,BoMaRa}.

To obtain this formulation for the scalar $K\pi$ form factor, one starts from a dispersion relation for the scalar polarization function of the $s$ and $u$ quarks \cite{BoMaRa,BC,AbAn,Hill}: 
\beq\label{eq:Pi0} 
\chi_{_0}(Q^2)\equiv \frac{\partial}{ \partial q^2} \left[ q^2\Pi_0 \right] 
= \frac{1}{\pi}\int_{t_+}^\infty\!dt\, \frac{t {\rm Im}\Pi_0(t)}{ (t+Q^2)^2} \,,
\eeq 
where unitarity implies the inequality: 
\beq\label{eq:unit}
{\rm Im} \Pi_0(t) \ge \frac{3}{2} \frac{t_+ t_-}{ 16\pi} 
\frac{[(t-t_+)(t-t_-)]^{1/2}}{ t^3} |f_0(t)|^2  \,,
\eeq
with $t_\pm=(M_K \pm M_\pi)^2$. We use here the notations from \cite{Hill}, where $\Pi_0$ is defined as the longitudinal part of the correlator of two vector currents. As in \cite{BC},  $\Pi_0$ can be identified with   $(\Psi(t)-\Psi(0))/t^2$, where $\Psi$ is the correlator of the divergence of the vector current.

 The quantity $\chi_{_0}(Q^2)$ in (\ref{eq:Pi0}) can be reliably calculated by pQCD when  $Q\gg\Lambda_{\rm  QCD}$. At present, calculations  available up to  the order $\alpha_s^4$  \cite{GeBr}-\cite{BaCh} give:
\bea\label{eq:pQCD}
\chi_0(Q^2)& =& \frac{3(m_s-m_u)^2}{8\pi^2 Q^2} \, \big[1 + 1.80\alpha_s  \nonumber \\
&+& 4.65\alpha_s^2 + 15.0\alpha_s^3 +57.4 \alpha_s^4 \dots\big], 
\eea
where the running quark masses and the strong coupling $\alpha_s$ are evaluated at the scale $Q^2$ in $\overline{MS}$ scheme.

Taking into account the fact that  $f_0(t)$ is analytic everywhere in the complex $t$-plane  except for the branch cut running from $\tplus$ to $\infty$, the relations (\ref{eq:Pi0})-(\ref{eq:pQCD}) can be expressed in the canonical form (\ref{eq:gI})  if one defines the variable
\beq\label{eq:z}
z(t)=\frac{\sqrt{t_+}-\sqrt{t_+-t}}{\sqrt{t_+}+\sqrt{t_+-t}},
\eeq
which maps the $t$ plane cut  from $\tplus$ to $\infty$ onto the unit disk $|z|<1$,  such that $z(0)=0$, and the function
\beq\label{eq:fg}
g(z)= f_0(t(z))\, w(z),
\eeq
where $t(z)$ is the inverse of (\ref{eq:z}) and $w(z)$ is the outer function \cite{BoMaRa}- \cite{AbAn}\footnote{We mention that a $\sqrt{2}$ is missing in the corresponding expressions given in \cite{BC,AbAn}.}: 
\bea	\label{eq:w}
 w(z)&=& \frac{\sqrt{3}}{32 \sqrt{\pi}} \,\frac{M_K-M_\pi}{M_K+M_\pi}\,(1-z)\,(1+z)^{3/2} \nonumber \\
&& \times\,\frac{ (1+z(-Q^2))^2} {(1-z\,z(-Q^2))^2}\frac{(1-z\, z(t_-))^{1/2}}{(1+ z(t_-))^{1/2}}. 
\eea
Then (\ref{eq:gI}) is satisfied, with 
\beq\label{eq:I}
I=\chi_0(Q^2).
\eeq
It may be noted that $z$ is an independent variable in the outer function,
whereas $z(-Q^2)$ etc.,
are defined {\it via} the conformal variable eq.(\ref{eq:z}). 

We use now the fact that, below the inelastic threshold $\tin$, the phase of the form factor $f_{0}(t)$  is known  from Watson's theorem and the $I=1/2$  $S$-wave  of elastic $K\pi$ scattering. Then one can define the Omn\`{e}s function
\beq	\label{eq:omnes}
 \omnes(t) = \exp \left(\D\frac {t} {\pi} \int^{\infty}_{\tplus} dt' \D\frac{\delta (t^\prime)} {t^\prime (t^\prime -t)}\right),
\eeq
where $\delta(t)$  is  the phase of the form factor known for $t\le \tin$, and is an arbitrary function, sufficiently  smooth ({\em i.e.} Lipschitz continuous) for $t>\tin$. It can be shown  \cite{AACIR2} that the results are independent of the function $\delta(t)$ for $t>\tin$.

Since the Omn\`{e}s function  $\omnes(t)$ fully accounts for the second Riemann sheet of the form factor, the  function $h(t)$, defined by
\beq	\label{eq:fh}
f_{0}(t) = h(t)\, \omnes(t),
\eeq
 is real analytic in the $t$-plane with a cut only for $t\ge \tin$. Then, the  relations (\ref{eq:Pi0})-(\ref{eq:pQCD}) and (\ref{eq:fh}) can be expressed in the canonical form (\ref{eq:gI}), by defining the new variable \cite{Caprini2000}
\beq\label{eq:z1}
z (t) = \D\frac {\sqrt{\tin}-\sqrt {\tin -t} } {\sqrt {\tin}+\sqrt {\tin -t}}, 
\eeq
which maps the $t$-plane cut for $t>\tin$  onto the unit disk $|z|<1$ of the $z$-plane, such that $z(0)=0$,   and define the function \cite{Caprini2000}
\beq\label{eq:fg1}
g(z) =  f_0(t(z))\,w(z)\, \omega(z) \,[\omnes(t(z))]^{-1},
\eeq 
where  $t(z)$ is now the inverse of (\ref{eq:z1}).
The new outer function $w(z)$  is defined as 
\bea	\label{eq:w1}
 w(z)&=& \frac{\sqrt{3}(M_K^2-M_\pi^2)}{16 \sqrt{2\pi} \tin}
 \frac{\sqrt{1-z}\,(1+z)^{3/2} (1+z(-Q^2))^2} {(1-z\,z(-Q^2))^2} \nonumber \\
&& \times\,\frac{(1-z\,z(t_+))^{1/2}\,(1-z\, z(t_-))^{1/2}}{(1+ z(t_+))^{1/2}\,  (1+ z(t_-))^{1/2}}, 
\eea
and 
\beq\label{eq:omega}
 \omega(z) =  \exp \left(\D\frac {\sqrt {\tin - t(z)}} {\pi} \int^{\infty}_{\tin} dt^\prime \D\frac {\ln |\omnes(t^\prime)|}
 {\sqrt {t^\prime - \tin} (t^\prime -t(z))} \right).
 	\eeq 
Then (\ref{eq:gI}) is satisfied, where $I$ is defined  as
\beq\label{eq:I1}
I= \chi_0(Q^2) - \frac{3}{2} \frac{t_+ t_-}{ 16\pi^2} \int_{t_+}^{\tin}\!dt\, \frac{[(t-t_+)(t-t_-)]^{1/2} |f_0(t)|^2}{t^2 (t+Q^2)^2}\,,
\eeq
and is calculable if the modulus $|f_0(t)|$ is known at low energies, below $\tin$.
Thus, we can use the inequality  (\ref{eq:det}) and the nonnegativity of the leading  minors  to obtain  bounds on the parameters of the expansion (\ref{eq:f0}).  The Taylor coefficients in (\ref{eq:gz}) are defined  uniquely in terms of these parameters by (\ref{eq:fg}) or  (\ref{eq:fg1}). We further choose $z_1=z(\Delta_{K\pi})$ and $z_2=z(\bar\Delta_{K\pi})$,  where $z$ is defined by (\ref{eq:z}) or (\ref{eq:z1}), and express $g(z_j)$ in terms of the values in (\ref{eq:CT}), by using  either (\ref{eq:fg}) or  (\ref{eq:fg1}). 

 In our analysis we take as inputs the values of the form factor at $t=0$ and $t=\Delta_{K\pi}$, the phase  below $\tin$ and the integral over the modulus required in (\ref{eq:I1}).  Then the constraints resulting from (\ref{eq:det}) restrict the coefficients $\lambda_0'$,  $\lambda_0''$ and the value of $f_0(t)$ at the second Callan-Treiman point $\bar\Delta_{K\pi}$.
\section{Input}\label{sec:input}
We  work in the isospin limit, adopting the convention that $M_K$ and $M_\pi$ are  the masses of the charged mesons. The inputs provided by the low energy theorems was discussed in the Introduction.

For choosing  $\tin$, we recall that the first inelastic threshold for the scalar form factor is set by the $K\eta$ state, which suggests to take $\tin=1\, \mbox{GeV}^2$ as in \cite{BC}. However, this channel has a weak effect, the elastic region extending practically up to the $K\eta'$ threshold,  which justifies the choice $\tin=(1.4\, \mbox{GeV})^2$. In our analysis we shall use for illustration these two values of $\tin$. 

Below $\tin$  the function $\delta(t)$ entering (\ref{eq:omnes}) is the phase of the $S$-wave of $I=1/2$ of the elastic $K\pi$ scattering  \cite{Jaminscalar2006,Bachirscalar2009}. In our calculations we use as default the phase from \cite{Bachirscalar2009}.  Above $\tin$ we assume $\delta(t)$ as a smooth function approaching $\pi$ at high energies. We  checked numerically that the bounds are independent of the choice of $\delta(t)$ for  $t>\tin$. 

 To estimate the integral appearing in (\ref{eq:I1}), we first used the  parametrization of $|f_0(t)|$  in terms of the resonances $\kappa$ and $K_0^*(1430)$, proposed by Belle collaboration  \cite{Belle}. Using as input the solution 1 in Table 4 and  Eq. (7)  of \cite{Belle}, the integral has  the value  $66.08 \times 10^{-6}$  for $\tin=1\, \mbox{GeV}^2$, and $184.89 \times 10^{-6}$  for  $\tin=(1.4\, \mbox{GeV})^2$. Although the parametrization used in \cite{Belle} does not have good analytic properties, this fact is not relevant for our analysis: all that we need is a numerical estimate of the integral in (\ref{eq:I1}).  The analyticity of the form factor is implemented rigorously in our approach, for every numerical input.

Aternatively,  using the modulus   available from the dispersive analyses \cite{Jaminscalar2006} or \cite{Bachirscalar2009}, the low energy integral in (\ref{eq:I1}) is $40.05 \times 10^{-6}$ or $37.01 \times 10^{-6}$, respectively, for $\tin=1\, \mbox{GeV}^2$, and  $89.31 \times 10^{-6}$ or $81.26 \times 10^{-6}$  for  $\tin=(1.4\, \mbox{GeV})^2$.

Finally,  we take $Q^2 = 4 \gev^2$  as in \cite{BC,Hill}, and obtain $\chi_0= (253 \pm  68 ) \times 10^{-6}$, using in (\ref{eq:pQCD}) $m_s(2 \gev)=98\pm 10 \mev,  m_u(2 \gev )=3\pm 1 \mev$ \cite{LL2009} and $\alpha_s(2 \gev)=0.308\pm 0.014$, which results from the recent average  $\alpha_s(m_\tau)=0.330\pm 0.014$ 
\cite{JaminBeneke,Davier,Bethke,CaFi2009}. 
The error of $\chi_0$ includes also a contribution of 15\%, of the order of magnitude of the last term in (\ref{eq:pQCD}), to account for the truncation of the expansion\footnote{In fact, it is known that the perturbative  series in QCD are divergent. Improved expansions exploiting this feature were proposed, see for instance \cite{CaFi2009}. However, in the present context this improvement is not necessary, since the sensitivity of the bounds to the value of $\chi_0$ is quite low.}.

\section{Results}\label{sec:results}
 In order to illustrate the effect of the additional information on the phase and modulus,  we compare in  Fig. \ref{fig:fig1}  the allowed domains in the plane $(\lambda_0',\, \lambda_0'')$, obtained with the standard and the new bounds, using only the constraint at $t=0$ (this case is  obtained from (\ref{eq:det}) by removing the lines and columns that contain $g(z_1)$ and  $g(z_2)$). The large ellipse is obtained with the standard bounds, (\ref{eq:z})-(\ref{eq:I}), the small ones represent the new bounds, calculated with  (\ref{eq:z1})-(\ref{eq:I1})  for two values of $\tin$.   For $f_0(0)$ we took the central value 0.962.
The  left panel is obtained with the integral in (\ref{eq:I}) calculated  with the modulus from \cite{Belle}, for the right one we used the modulus from  \cite{Jaminscalar2006}.

 The inner ellipses are slightly smaller  in the left panel than in the right one, because  in the latter case the  integral in (\ref{eq:I1}) is smaller  and  $I$ is larger (it is easy to see that a larger value of $I$ leads to an ellipse of a larger size). Moreover, in the right panel the small ellipses are not contained entirely inside the large one, which means that among the  functions satisfying the constraints  (\ref{eq:z1})-(\ref{eq:I1}) there are some that violate the original  bounds  (\ref{eq:z})-(\ref{eq:I}).  However, this does not mean that the conditions are inconsistent, since the ellipses have a nonzero intersection, which represents the allowed domain in this case.

\bfig[ht]
	\begin{center}\vspace{0.cm}
	 \includegraphics[angle = 0, clip = true, width = 2.3 in]
{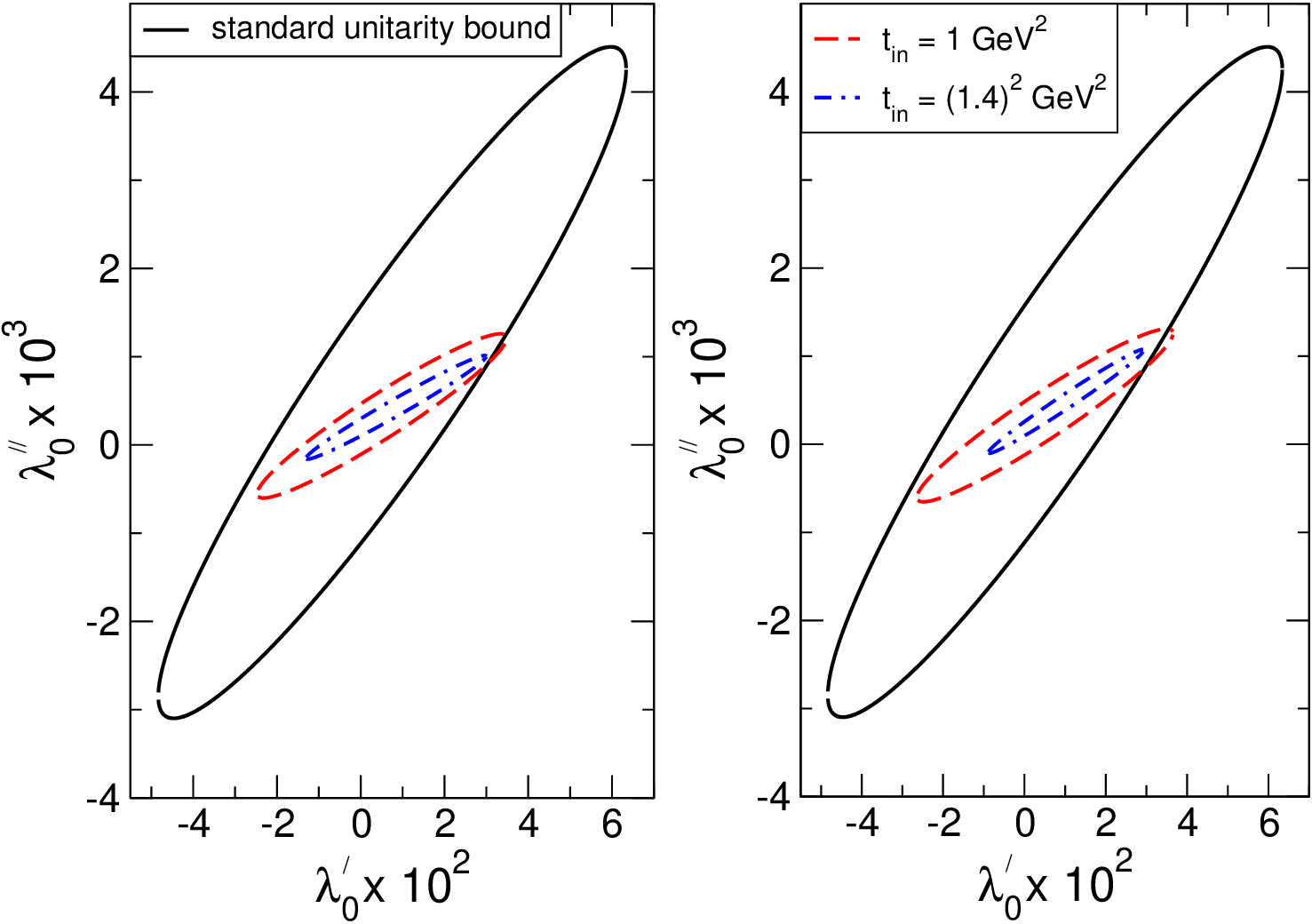}
	\end{center}\vspace{-0.2cm}
	\caption{Allowed regions for the slope and curvature from the standard 
and the new bounds with the constraint  at $t=0$. Left: modulus from \cite{Belle}; Right: modulus from \cite{Jaminscalar2006}.}
	\label{fig:fig1}
\efig

The effect of the low-energy theorems (\ref{eq:CT}) is illustrated in Fig. \ref{fig:fig2}, which shows the improved bounds (\ref{eq:z1})-(\ref{eq:I1}) calculated with  the modulus from \cite{Belle} for $\tin=1\,\gev^2$.  The large ellipse is obtained using the constraints at $t=0$ and $t=\ct$, the tiny ellipses result from the constraints at $t=0$,  $\ct$ and $\ctbar$, for  the central values of $f_0(0)$ and $f_0(\ct)$ and several values of the  correction $\bar\Delta_{CT}$. The strong constraining power of the simultaneous constraints at $\ct$ and $\ctbar$ was noted in \cite{AbAn}. However, the bounds derived now are much stronger than those in \cite{AbAn}, due to the additional  information on the phase and modulus on the cut.

\bfig[ht]	
	\begin{center}\vspace{0.cm}
	 \includegraphics[angle = 0, clip = true, width = 2.3 in]
{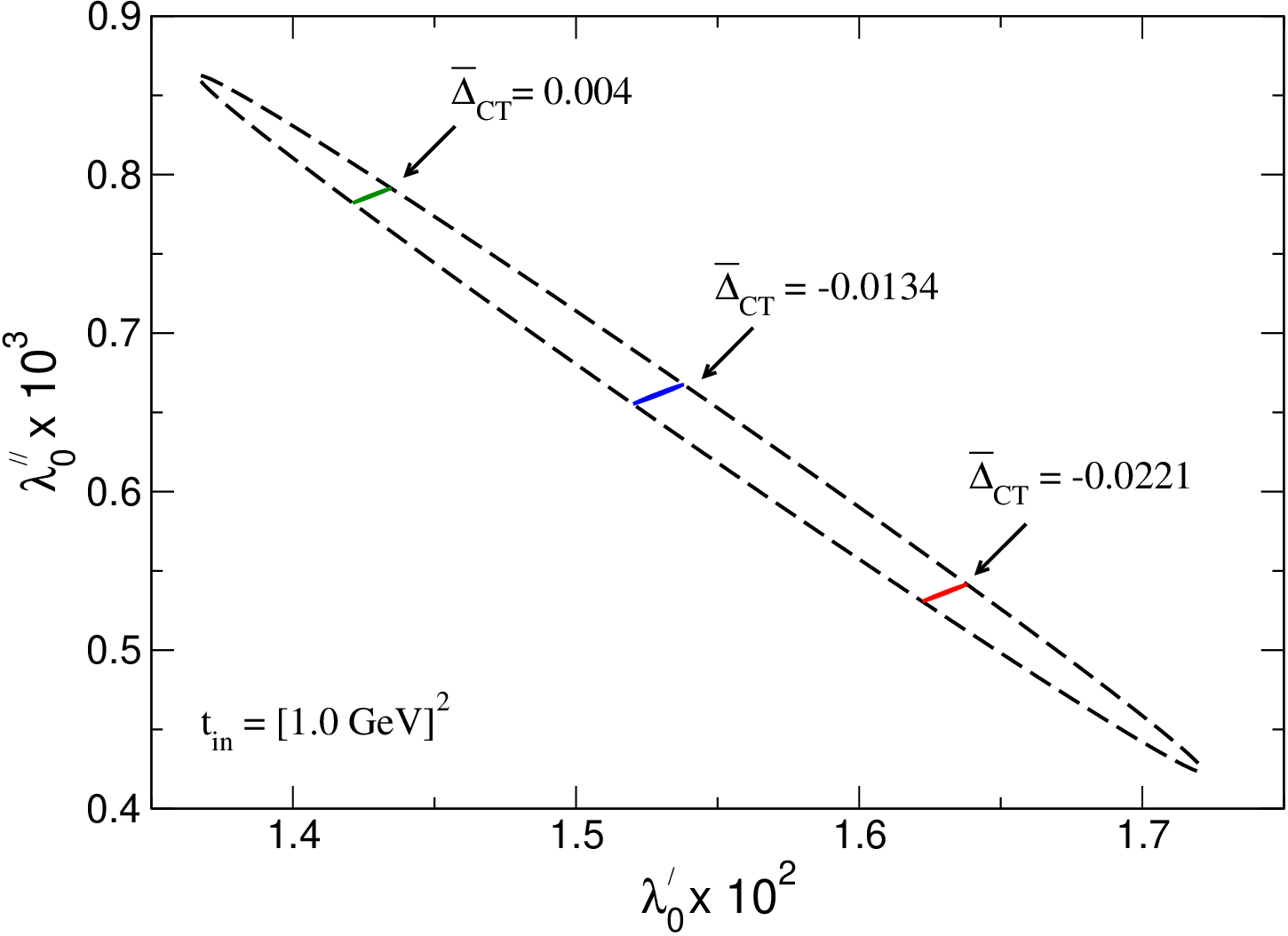}	
	\end{center}\vspace{-0.2cm}
	\caption{Large ellipse:  new bounds obtained with  $\tin=1\,\gev^2$
 and the values at $t=0$ and $\ct$; small ellipses: new bounds for $\tin=1\,\gev^2$ and simultaneous constraints at $t=0$, $\ct$ and $\ctbar$, for several  values of $\bar\Delta_{CT}$.}  
	\label{fig:fig2}
\efig

 The small ellipses exist only for $\bar\Delta_{CT}$ inside a rather narrow interval, whose end points  lead to  inner ellipses of zero size in Fig. \ref{fig:fig2}.  Actually, this range results directly  from the inequality  (\ref{eq:det}): by keeping only the lines and columns involving $g_0$, $g(z_1)$ and $g(z_2)$ and using the central values at $t=0$ and $\ct$, we obtain  $-0.046 \le   \bar\Delta_{CT} \le  0.014$.

We recall that  the current ChPT prediction is  $-0.057 <  \bar\Delta_{CT} < 0.089$ (cf. Eq. (4.9) of \cite{BePa} adapted to our input  $F_K/F_\pi$). As this interval is larger  than the range derived above, we conclude that, at present,  one can not  further restrict the domain for the slope and curvature  using  the low-energy theorem at the second Callan-Treiman point: by varying 
$\bar\Delta_{CT}$ inside its currently  known range we obtain the union of the tiny  ellipses in Fig. \ref{fig:fig2}, which covers the large ellipse obtained using only the value at $\ct$.

 In Fig. \ref{fig:fig3} we show the allowed domains  for the slope and curvature obtained with the constraints at $t=0$ and $\ct$  for  two values  of $\tin$
in the left panel. In the right panel we also superimpose the allowed region
when no phase and modulus information is taken into account.  
As in Fig. \ref{fig:fig2},  the low energy integral in (\ref{eq:I1}) was calculated with the modulus from \cite{Belle}.
 For  $\tin = 1\, \gev^2$  the large ellipse implies the  range $0.0137 \le \lambda_0'\le 0.0172$, for  $\tin = (1.4\, \gev)^2$ the small ellipse  implies the narrower range  $0.0150 \le \lambda_0'\le 0.0163$. In both cases we obtain  a strong  correlation between the slope and the curvature. 
It may be clearly seen that a dramatic improvement is obtained by the
inclusion of phase and modulus data.
 
\bfig[ht]\vspace{-0.cm}
	\begin{center}
	 \includegraphics[angle = 0, width = 2.3 in, clip = true]{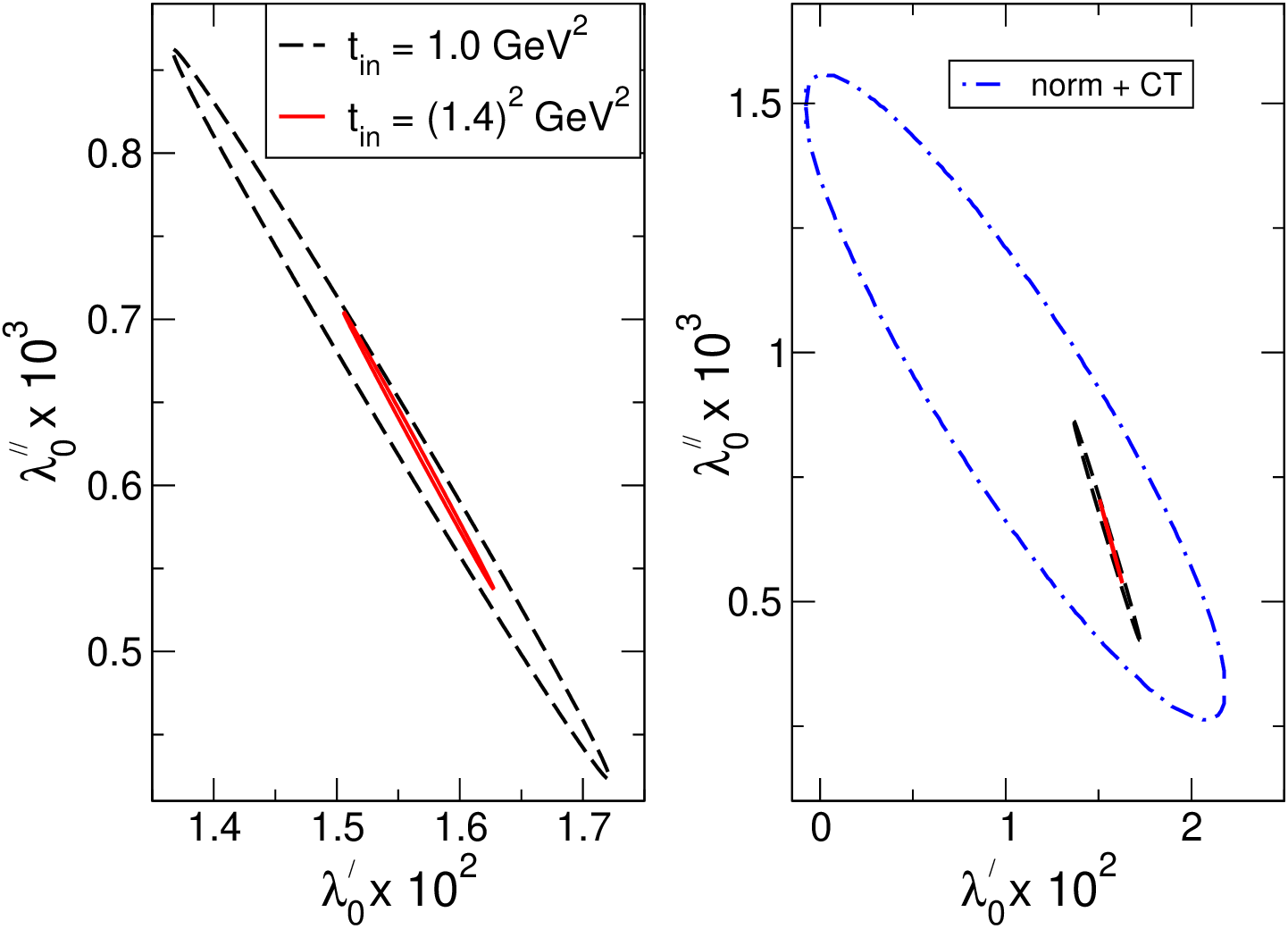}
	\end{center}\vspace{-0.2cm}
	\caption{Left:
Allowed regions for the slope and curvature using as input 
the phase and modulus up to $\tin$, and the values of $f_0(0)$ and  $f_0(\ct)$;
right: as in left panel and also showing the region obtained with no phase
and modulus information. }
	\label{fig:fig3}
\efig

As discussed, the method gives also very sharp predictions for the corrections  $\bar\Delta_{CT}$ at the second Callan-Treiman point:  
for  $\tin = (1.4\, \gev)^2$ we obtain
$-0.031 \le   \bar\Delta_{CT} \le  -0.008$ 

The above ranges were obtained using the central values of $f_0(0), \,f_0(\ct)$ and $\chi_0$, the phase from \cite{Bachirscalar2009} and the modulus from \cite{Belle}.  Accounting for the errors  and using alternatively the phase and modulus  from \cite{Jaminscalar2006},  the end points of the  range of  
$\lambda_0'$   for  $\tin = (1.4\, \gev)^2$ varied by $ \pm 0.00039|_{0} 
\pm 0.00044|_{\ct} \pm 0.00028|_{\chi_0}\pm 0.00038|_{\rm mod} \pm 
0.00070|_{\rm ph}$, while for   $\bar\Delta_{CT}$ the variation was  
$\pm 0.0074|_{0} \pm  0.0002|_{\ct} \pm 0.0052|_{\chi_0} \pm 0.0069|_{\rm mod} 
\pm 0.0020|_{\rm ph}$.  
We note that, while the bounds are very sensitive to the input value of 
$f_0(0)$,  the uncertainty  of $\chi_0$ has a relatively low influence on 
the results.

In conclusion, our analysis shows that the modified type of unitarity bounds proposed in \cite{Caprini2000}, which includes input from the elastic part of the  cut, leads to very stringent  bounds on the scalar  $K\pi$ form factor  at low energy.   Using as input the precise values at $t=0$ and $\ct$ and assuming 
that the inelasticity is negligible  below (1.4 GeV)$^2$, 
we obtain for the slope
at $t=0$ the range  $0.0150(10) \le \lambda_0'\le 0.0163(10)$, where the error is obtained by adding in 
quadrature the uncertainties due to various  inputs.  As shown in  Fig. \ref{fig:fig3}, the method leads to a strong correlation  between the slope and the curvature. 
We obtain also a narrow admissible range 
$-0.031(12)\leq \bar{\Delta}_{CT}\leq -0.008(12)$
for the higher order ChPT corrections  at the second 
Callan-Treiman point $\ctbar$, significantly reducing the range from ChPT mentioned earlier. Unlike in the usual dispersive  approaches, the predictions are independent of any assumptions about the 
presence or absence  of zeros, or the phase and modulus of the form factor above the inelastic  threshold.


\begin{acknowledgement}
We are grateful to  H. Leutwyler for useful suggestions and to 
B. Moussallam and M. Jamin for supplying us their parametrizations 
of the $K\pi$ form factors. BA thanks DST, Government of India,  
and the Homi Bhabha Fellowships Council for support. 
IC acknowledges support from  CNCSIS in the Program Idei, Contract No.
464/2009, and from ANCS, project PN 09 37 01 02 of IFIN-HH.
\end{acknowledgement}



\end{document}